\documentclass[preprint,12pt,authoryear]{elsarticle}

\usepackage{amssymb}
\usepackage{amsmath}

\usepackage{multirow}
\usepackage{float}
\usepackage{hyperref}
\usepackage{tabularx}
\usepackage{booktabs}

\journal{International Journal of Human-Computer Studies}

\begin{document}

\begin{frontmatter}



\title{Assessing the Applicability of Existing Design Recommendations to AI Companion Design: A Multi-Method Study\tnoteref{cc}} 


\author[inst1]{Soobin Cho\corref{cor}}
\ead{soobin30@uw.edu}

\author[inst1]{Deveshi Modi\fnref{equal}}
\author[inst1]{Divya Mavinkurve\fnref{equal}}
\author[inst1]{Jieqiong Ding\fnref{equal}}
\author[inst1]{Mark Zachry}

\affiliation[inst1]{organization={Human Centered Design \& Engineering, University of Washington},
            addressline={1410 NE Campus Pkwy}, 
            city={Seattle},
            postcode={98195}, 
            state={Washington},
            country={United States}}

\cortext[cor]{Corresponding author.}
\fntext[equal]{These authors contributed equally to this work.}
\tnotetext[cc]{\copyright~2026. This manuscript version is made available under the CC BY-NC-ND 4.0 license \url{https://creativecommons.org/licenses/by-nc-nd/4.0/}. This is the accepted manuscript of an article published in the \textit{International Journal of Human-Computer Studies}. The formal publication is available at \url{https://doi.org/10.1016/j.ijhcs.2026.103940}.}

\begin{abstract}
With the rapid proliferation of large language model (LLM)-based systems, AI companions have emerged as conversational agents designed to cultivate emotional connection rather than primarily to support humans in instrumental tasks. Because engagement with AI companions involves relational, emotional, and potentially long-term interactions, their design is consequential. Prior work has offered guidance for designing trustworthy and relational AI systems and has begun to examine design for AI companionship. However, while such work provides insights into possible design solutions, less is known about what makes AI companion design difficult as a design problem. To examine this challenge, we assessed the applicability of existing design recommendations from adjacent domains in the context of AI companion design. Our multi-method investigation unfolded across four phases: literature review, practitioner co-analysis, internal heuristic evaluation, and external expert assessment. Throughout this process, we synthesized nine design principle areas that surfaced tensions in the applicability of existing recommendations to AI companion design. Our findings show that ethical and UX-oriented considerations are deeply intertwined and often require context-sensitive application. We document a systematic, multi-method problem analysis that uses these principle areas as an analytic artifact to examine why existing recommendations cannot be directly transferred to AI companion contexts.
\end{abstract}


\begin{highlights}
\item A multi-method study assessed design recommendations for AI companions
\item Nine design principle areas were identified
\item The principle areas surfaced tensions in recommendation applicability
\item Ethics and user experience considerations are deeply intertwined in AI companions
\item Effective AI companion design requires context-sensitive flexibility
\end{highlights}


\end{frontmatter}



\section{Introduction}

Large language models (LLMs) have contributed to the rise of AI companions, a distinct type of conversational system designed to foster emotional connection, expressive conversation, and ongoing interaction. Rather than focusing primarily on task completion, AI companions prioritize relational and affective engagement, often adopting human-like personas to build intimacy~\citep{de2026ai, Weitzman}.

Recent research has increasingly examined user experiences of AI companions. Studies have examined how users develop and experience relationships with AI companions, including attachment, trust, love, and interpersonal connections~\citep{ng2026love, silayach2025algorithms, sharpe2024exploring, merrill2022ai, manoli2026digital}. Studies have also examined psychosocial outcomes associated with AI companion use, including the use of AI companions for emotional support, their potential to reduce loneliness, their promotion of self-reflection, and their influence on user well-being~\citep{zhang2025rise, ng2026love, de2026ai, kouros2024digital, Vivian, Li, Maples, Skjuve}. At the same time, ethical concerns have been raised surrounding AI companionship, including  emotional harm, emotional manipulation, harmful dependency, erosion of human relationships, exploitation of loneliness, perpetuation of gender and racial stereotypes, and problematic sexual interactions~\citep{boine2023emotional, brigham2026examining, muldoon2025cruel, ho2025potential, Banks, Skjuve, Verma, Jacobs}. Together, this growing body of work underscores the importance and complexity of AI companion design.

Prior work has offered design guidance relevant to AI companions, including broader guidance for trustworthy AI design, research on relational and anthropomorphic systems, and work that more directly examines AI companion design. However, while such work provides valuable design guidance and proposed solutions, less attention has been paid to understanding why AI companion design remains difficult as a design problem.

To systematically examine the challenges of AI companion design, we assessed how recommendations from adjacent domains apply to AI companions. To do this, we conducted a multi-method investigation involving four phases: Phase 1 collected design recommendations through a comprehensive literature review; Phase 2 assessed these recommendations for AI companion design contexts through collaborative analysis with 11 practitioners; Phase 3 integrated practitioner inputs into cohesive design principles and assessed them through internal heuristic evaluation; and Phase 4 conducted a community check of these principles through modified heuristic evaluation with 14 domain experts.

This multi-method problem analysis identified nine design principle areas relevant to AI companions: Safety, Transparency, Inclusivity, Predictability and Consistency, Interaction Controllability, Adaptation and Personalization, Engagement, Empathetic Response, and Response Length. Across these areas, we found tensions in applying existing recommendations to AI companion design. We report two key findings: (1) ethical and UX-oriented considerations are deeply intertwined in AI companion design, with many principles shaping both experience and risk, and (2) principles often require context-sensitive application rather than fixed, one-size-fits-all guidelines.

In this paper, we systematically document this multi-method problem analysis, using the principle areas as an analytic artifact.

\section{Related Work}
\subsection{Designing Trustworthy AI}
Prior work has moved trustworthy AI from abstract principles to lifecycle-based implementation. \citet{LiBo} organize trustworthiness requirements such as robustness, explainability, transparency, reproducibility, fairness, privacy, and accountability across the AI system lifecycle, from data and model development to deployment, monitoring, and governance. \citet{Toreini} similarly connect trust in AI to trustworthy machine learning technologies, identifying fairness, explainability, auditability, and safety as technologies that support trust across the lifecycle of AI-based systems.

Human-centered AI (HCAI) research shifts this discussion from automation-centered AI toward human-centered trustworthy design. \citet{shneiderman2020human} proposes a two-dimensional HCAI framework in which reliable, safe, and trustworthy AI systems combine high levels of automation with high levels of human control, supported by mechanisms such as audit trails, benchmark testing, informative feedback, reversible actions, and independent oversight. \citet{Schoenherr} also treats explainability and accuracy as important for trustworthiness, while embedding them in broader design processes involving multi-stakeholder engagement, interdisciplinary knowledge, and socio-technical context. This line of work positions trustworthy AI design as a matter of preserving meaningful human agency within automated systems.

Human-AI interaction (HAI) and human-computer interaction (HCI) work further shifts attention from trustworthy system properties to how users encounter and calibrate trust through interaction. \citet{LiaoSundar} conceptualize trustworthiness as communicated through cues embedded in AI-generated content, transparency features, and interaction affordances, which users interpret through systematic or heuristic processing. This perspective shows that transparency and explanation do not automatically produce appropriate trust, and may instead lead to over-trust, over-reliance, or inappropriate trust.

Finally, assessment-oriented work moves from general guidance to situated socio-technical assessment. \citet{Vetter} note that many trustworthy AI guidelines remain high-level and abstract, making it difficult to assess whether a specific AI system satisfies them in practice. Their work on Z-Inspection shows how trustworthy AI guidance can be operationalized through socio-technical scenarios, stakeholder analysis, lifecycle-based assessment, and the identification of ethical tensions.

Taken together, prior work has translated trustworthiness principles into design-oriented guidance for lifecycle practices, human-centered control, interactional trust calibration, and situated socio-technical assessment.

\subsection{Designing Relational and Anthropomorphic Systems}
Prior work on relational and anthropomorphic systems has examined how conversational agents, embodied conversational agents, and social robots can be designed to appear more human-like, social, or relational.

Anthropomorphic conversational agent research specifies how perceived anthropomorphism can be shaped through design. In their design framework, \citet{Seeger1} identify human identity cues, verbal cues, and nonverbal cues as key dimensions for designing anthropomorphic conversational agents. Their later work further shows that perceived anthropomorphism does not simply increase as more human-like cues are added; instead, cue combinations, task type, and users’ disposition to anthropomorphize can affect how anthropomorphic design cues are perceived~\citep{Seeger2}.

Embodied conversational agent research frames face-to-face conversational interaction as a problem of coordinating verbal and nonverbal behavior, rather than merely giving agents human-like bodies. \citet{Cassell1} define embodied conversational agents as agents that can recognize and produce verbal and nonverbal input and output, handle turn-taking, feedback, and repair, and signal the state of the conversation. \citet{Cassell2}’s account of face-to-face conversation for embodied conversational agents further specifies design-relevant behaviors such as gesture, gaze, facial displays, timing, and conversational functions.

Personality and empathy have also been treated as design targets in conversational agent and social robot research. \citet{Ahmad} organize conversational agent personality cues across verbal and nonverbal expression, and \citet{Dregger} review language cues used to express artificial personality. \citet{Volkel} show that the Big Five model may not fully capture how users describe speech-based agent personality, motivating agent-specific personality models. In social robot research, \citet{Park} frame empathy as a design problem involving the recognition of users’ emotional states, thoughts, and situations, affective or cognitive response generation, empathy modulation, and multimodal expression.

\subsection{Designing AI Companions}
Compared with broader work on the effects, risks, and user experiences of AI companions, design-focused research remains relatively limited.

One of the most explicit attempts to theorize AI companion design is \citet{Strohmann}’s design theory for virtual companionship. They develop prescriptive design knowledge for conversational agents designed to support companionship, proposing design principles around human-like design, adaptivity and adaptability, proactivity and reciprocity, transparency, privacy, ethics, and relationship.

Other design-focused work examines AI companions in specific support contexts. In informal caregiving, an AI-driven chatbot was designed to help caregivers recognize their role, reflect on their situation, and identify relevant support options, leading to design principles around personalized assessment, transparent information, role awareness support, accessibility, and continuous companionship~\citep{Walter}. In foreign language writing, an interactive AI companion was designed to support a process-oriented approach to writing that emphasizes learner autonomy, collaboration, and textual conventions~\citep{Mills}.

\vspace{1em}
Taken together, prior work suggests that AI companion design can draw on trustworthy AI guidance, relational and anthropomorphic systems research, and emerging design guidance for AI companions. However, these bodies of work mainly specify what designers should consider or aim to achieve, rather than examining the difficulties that arise when such guidance is applied to AI companion design. As a result, researchers have a limited understanding of why AI companion design remains difficult as a design problem. To address this gap, we examined the applicability of existing recommendations in the context of AI companion design.

\section{Overview of the Multi-Method Study}
\begin{figure}[!htbp]
    \centering
    \includegraphics[width=1\linewidth]{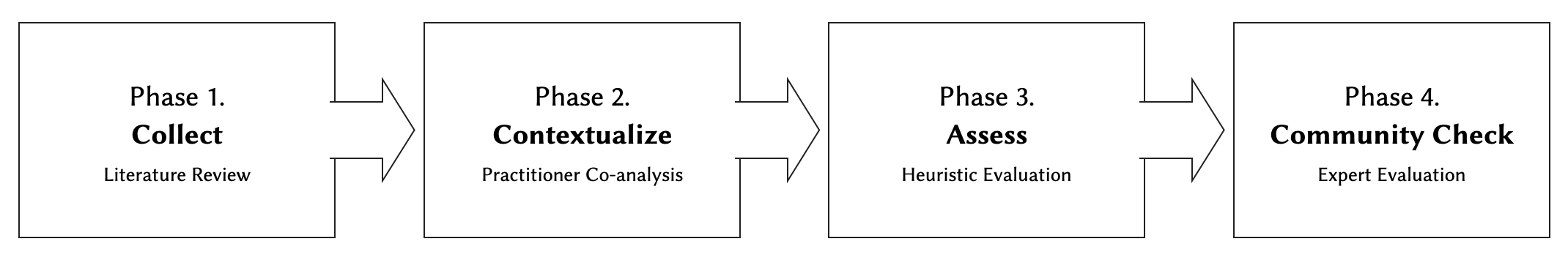}
    \caption{Multi-method study design illustrating the progressive assessment of design recommendations across four phases: Collect (literature review), Contextualize (practitioner co-analysis), Assess (internal heuristic evaluation), and Community Check (expert evaluation).}
    \label{fig:method}
\end{figure}

To examine how design recommendations from adjacent domains apply to AI companions, we conducted a four-phase multi-method problem analysis. Rather than treating the phases as separate studies, we structured them as a progressive assessment of recommendation applicability. This phase structure was informed by prior heuristic evaluation approaches, with each phase producing an intermediate output that informed the next phase (Figure~\ref{fig:method}).

In Phase (1) \textbf{Collect}, we built the initial corpus of existing design recommendations. Through a literature review and filtering process, we identified 548 recommendations from 41 published sources that were potentially relevant to AI companion design.

In Phase (2) \textbf{Contextualize}, we worked with 11 experienced practitioners to assess how the collected recommendations applied to AI companion use. Through collaborative coding and discussion, we examined the recommendations in context and identified initial applicability tensions.

In Phase (3) \textbf{Assess}, we synthesized the applicable recommendations and practitioner insights into an initial set of design principles and conducted an internal heuristic evaluation with two AI companion products. This phase helped refine the principles and organize them into nine design principle areas, while also providing an initial assessment of how they applied in actual AI companion products.

Finally, in Phase (4) \textbf{Community Check}, we conducted a modified heuristic evaluation with 14 domain experts. This phase examined how the refined principles from Phase 3 applied across dedicated AI companion products and general-purpose conversational AI products used in companion-like contexts.

\section{Phase (1) Collect: Literature Review}

\subsection{Procedure}
Phase 1 aimed to build an initial corpus of existing design recommendations for assessing their applicability to AI companion design in later phases. Because our goal was not to conduct an exhaustive systematic literature review, we used a purposive, recommendation-oriented literature review approach. We focused on sources that explicitly articulated design principles, guidelines, recommendations, or design-relevant findings that could inform high-level user experience design for AI companions. To capture recommendations from adjacent domains, we organized the review around four categories: 1) General HCI/UX, 2) Human-AI interaction, 3) Agent and chatbot, and 4) AI companion.

We applied initial filtering both at the document level and the individual recommendation level to retain only sources and recommendations relevant to high-level user experience design. Specifically, we excluded recommendations that: addressed general attitudes beyond product design (e.g., always be cautious, how to assemble a team); focused on visual UI elements (e.g., menu layout, font); involved technical implementation (e.g., model tuning); or were overly tailored to specific domains or brands (e.g., marketing, healthcare).

For the \textbf{General HCI/UX} category, we sought foundational and widely used usability, interaction design, and user experience guidance. We identified classic HCI/UX principles and guidelines through targeted searches in Google Scholar and Google Search, and retained sources that provided explicit, generalizable recommendations for designing user interactions. We also purposively reviewed publicly available design ethics, privacy, and inclusion guidance from major technology companies, retaining those that provided generalizable, practitioner-facing design recommendations.

For the \textbf{Human-AI interaction} category, we sought design guidance on how AI systems should interact with users, with attention to both interactional and ethical considerations. We purposively reviewed publicly available AI design and responsible AI guidance from major technology companies, retaining documents that provided generalizable, practitioner-facing recommendations. We also searched the ACM Digital Library using keywords such as ``Human-AI interaction,'' ``AI,'' ``AI ethics,'' and ``ethical AI,'' combined with ``guidelines'' or ``principles,'' and retained academic sources that provided explicit design principles or frameworks relevant to human-AI interaction.

For the \textbf{Agent and chatbot} category, we first considered early, general recommendations that were developed when agents were first conceptualized as intelligent machines capable of human-like interaction and task automation~\citep{NormanAgent, Soltysiak}. We identified classic agent-related sources through targeted searches in Google Scholar and Google Search using the keyword ``agent.'' We also reviewed more recent design guidance for conversational agents and chatbots. For practitioner-facing guidance, we purposively reviewed publicly available conversation design and responsible bot guidance from major technology companies. For academic sources, we searched the ACM Digital Library and Google Scholar using keywords such as ``conversational agent'' and ``chatbot,'' combined with ``guidelines'' or ``principles.'' In conversational agent documents, we focused on conversation design recommendations and excluded those related to UI design (e.g., button layout), implementation details (e.g., model training), multimodality (e.g., camera/microphone integration), or assistive task-specific guidance (e.g., authentication systems).

For the \textbf{AI companion} category, we reviewed publications on both companion machines and AI companions. First, we searched the ACM Digital Library and Google Scholar using keywords such as ``companion'' and ``robot,'' combined with ``guidelines'' or ``principles,'' to identify design guidance for other types of companion machines, such as non-conversational robots. We then searched the ACM Digital Library, Google Scholar, and Google Search using ``AI companion'' combined with ``guidelines'' or ``principles'' to identify sources specifically focused on AI companion design. Because AI companion-specific design guidelines remain limited, we also drew on user studies that examined relational factors shaping intimacy, trust, social support, and companionship between users and AI companions. These studies were retained when their findings offered design-relevant insights for AI companion interactions.

\subsection{Output}
This process resulted in 548 retained design recommendations, including principles, guidelines, and user study findings, from 41 final retained sources. These sources were the documents from which the retained recommendations were extracted, rather than the full set of candidate sources considered during the search and screening process. The final sources included nine related to General HCI/UX, 11 to Human-AI interaction, 13 to Agent and chatbot, and eight to AI companion (Table~\ref{tab:LitReview}).

\begin{table}[!htbp]
\centering
{\small
\begin{tabular}{|p{3cm}|p{6cm}|c|c|}
\hline
\multicolumn{1}{|c|}{\textbf{Category}} & 
\multicolumn{1}{c|}{\textbf{Retained Sources}} & 
\textbf{\shortstack{\# of \\Sources}} &
\textbf{\shortstack{\# of \\Recs.}} \\
\hline

General HCI/UX
& \citep{Apple1987, Sydney, Rolf, NielsenBook, Nielsen_1994, Norman, ApplePrivacy, AppleInclusion, AppleWriting} 
& 9 & 101 \\ \hline

Human-AI interaction
& \citep{AppleDev, GooglePAIR, SAPgen, Amershi, Weisz, GoogleAI, IBMpillars, IBMethics, SAPtrustgen, Mohseni, Jakesch} 
& 11 & 229 \\ \hline
 
Agent and chatbot
& \citep{NormanAgent, Horvitz, Soltysiak, Carley, AZconv, AZnat, AZerr, MSbot, MSbot10, Stieglitz, Radziwill, Langevin, Silva} 
& 13 & 161 \\ \hline

AI companion
& \citep{Banks, Hoffman, Irfan, Strohmann, Khosrawi-Rad, Leo-Liu, Skjuve2, Vivian} 
& 8 & 57 \\ \hline

\textbf{Total} 
& 
& \textbf{41} & \textbf{548} \\ \hline

\end{tabular}
}
\caption{Categories, retained sources, and number of retained sources and design recommendations.}
\label{tab:LitReview}
\end{table}

\section{Phase (2) Contextualize: Practitioner Co-analysis}
With a team of 11 HCI practitioners, we engaged in collaborative coding and group discussions of all 548 recommendations to translate them into the context of AI companions.

\subsection{Participants}
We recruited the team participants from an academic institution located in the western United States. During recruitment, we screened for individuals with at least six months of full-time or internship experience in the industry. A total of 11 participants (9 female, 2 male) formed our research team in this phase. The team brought diverse combinations of HCI-related experience: one had a mix of UX research, design, and product management; one had experience across design, engineering, and product management; one had both UX research and design experience; and one had both design and engineering experience. In addition, two participants specialized in UX research, three in design, one in engineering, and one in product management. The team’s overall experience in HCI/UX ranged from 6 months to over 10 years: five had 6 months to 1 year of experience, four had 3 to 5 years, one had 5 to 7 years, and one had more than 10 years.

\subsection{Procedure}
This phase was conducted over ten weekly meetings, spanning more than two months. The first four meetings were designed to help team participants become familiar with AI companion products. During this period, each participant tried three to five different AI companions available in the market and shared their personal experiences during group discussions. To quickly gain user perspectives, participants also engaged in lightweight user research. Five participants conducted guided conversations with an actual AI companion user from their personal networks, asking about their usage history, motivations, conversational experiences, emotional responses, perceived benefits and drawbacks, and how their perceptions and behaviors evolved over time. The remaining six participants conducted online observations of user contributions on the Replika Reddit server, each selecting a different time range within the past year. They recorded notable posts, capturing the date, title, content, images, number of comments, and number of upvotes. They also summarized key discussions in the comments and took field notes on insights they found noteworthy. Team participants shared their findings and discussed their insights during the weekly meetings.

In the remaining six meetings, participants collaboratively evaluated the applicability of the 548 design recommendations to user experience design of AI companions. The recommendations were divided in half, each individually coded by five to six team members along with notes explaining their reasoning. These individual analyses were followed by collective group discussions, in which we examined each recommendation in turn, assessing its applicability and translating it into the context of AI companions.

\subsection{Results}
Of the 548 design recommendations reviewed, 95 were deemed inapplicable to AI companion design, while 453 were considered applicable.

Among the inapplicable recommendations were those emphasizing efficient use and direct modification. For example, principles such as \emph{Efficiency of use}~\citep{NielsenBook, Nielsen_1994, Langevin}, \emph{Minimal user actions}~\citep{Sydney}, and \emph{Provide shortcuts}~\citep{Rolf} were rejected because conversational interaction itself is the goal in AI companions, and making such systems  more efficient compromises that goal. Recommendations related to directly modifying unsatisfactory outputs, whether by canceling the output (e.g., \emph{Actions should be reversible}~\citep{Norman}) or editing it (e.g., \emph{Support co-editing of generated outputs}~\citep{Weisz}), were rejected. Since AI companions prioritize relationship-building through conversation, revising prior messages was seen as unnatural, and recovery through continued dialogue was preferred.

Among the applicable recommendations, we found that principle areas often treated as ethics-oriented, such as Transparency and Inclusivity, may significantly impact user experience, suggesting that ethical and UX considerations can be closely connected in AI companion design. Additionally, while principle areas such as Safety, Predictability and Consistency, Adaptation and Personalization, Engagement, and Empathetic Response should generally be upheld, we found that Transparency, Inclusivity, Predictability and Consistency, Interaction Controllability, and Response Length may require context-sensitive application for optimization of the user experience.

\subsubsection{Transparency}
We identified four types of transparency: Identity Transparency, Capability and Limitation Transparency, Information Transparency, and Decision Transparency. We agreed that all should be disclosed, but discussed that the frequency of disclosure should remain flexible. For Identity Transparency, although identifying the AI (e.g., \emph{Self-identify the agent as a machine}~\citep{Stieglitz}) is important, we noted that repeatedly stating this throughout a conversation can harm the relational experience essential for building companionship. We discussed similar concerns for Information Transparency (e.g., \emph{We cite the data sources [...]}~\citep{SAPgen}, \emph{Provide a confidence rating or uncertainty indicator}~\citep{SAPtrustgen}) and Decision Transparency (e.g., \emph{Make clear why the system did what it did}~\citep{Amershi}), where repeating such disclosures too often may harm the desired user experience. Instead, we suggested communicating them only when factual accuracy or the system’s reasoning is especially relevant during ongoing conversation.

\subsubsection{Inclusivity}
Among inclusivity-related recommendations, we noted that the recommendation to \emph{Use diverse names as examples}~\citep{AppleWriting} may not suit users who prefer content aligned with a specific region or culture. The recommendation to \emph{Consider carefully before including humor}~\citep{AppleInclusion} was also seen as requiring contextual adjustment to avoid excessive censoring that could diminish a desired sense of fun in interactions with AI companions. Recommendations encouraging the use of plain language, such as \emph{Avoid technical or statistical jargon}~\citep{AppleDev} and \emph{Replace colloquial expressions with plain language}~\citep{AppleInclusion}, were understood to depend on users’ familiarity with the domain and should be adjusted accordingly.

\subsubsection{Predictability and Consistency}
We noted that while highly unpredictable system behavior is undesirable, overly consistent and predictable behavior can diminish the enjoyment of AI companion use. Accordingly, for recommendations emphasizing consistency such as \emph{Consistency}~\citep{Apple1987} and \emph{Consistency of the system}~\citep{Norman}, we concluded that AI companions do not have to be strictly consistent. We were more cautious about predictability. The recommendation \emph{Predictability: Intelligibility supports building a mental model of the system that enables users to predict system behavior}~\citep{Mohseni} was rejected following discussion, as we viewed some degree of unpredictability as important for open-ended, companion-like conversation. In addition, when discussing whether \emph{Symbiotic agency} described by \citet{Leo-Liu} is applicable to AI companions, we considered their view that AI-simulated intimacy does not guarantee full controllability or predictability and may involve uncertainties similar to human relationships, and therefore determined this element to be applicable.

\subsubsection{Interaction Controllability}
While many existing recommendations emphasized user control~\citep{Apple1987, NielsenBook, Nielsen_1994, GooglePAIR, GoogleAI, Weisz, SAPtrustgen, Langevin}, we discussed that too much control may reduce engagement in the context of AI companions. Instances where the companion initiates a conversation or attempts to continue one that seems to be winding down were noted as potentially enjoyable. While core system-level actions like turning the app on and off should remain under user control, we saw potential value in these AI companion-led behaviors to support playful engagement.

\subsubsection{Response Length}
While prior design recommendations on response length tend to emphasize minimal and concise responses (e.g., \emph{Use the fewest words to convey the most meaning}, \emph{Surface only relevant messages to the customer}, \emph{Don’t state the obvious}~\citep{AZnat, AZconv}), we noted that these recommendations are not uniformly applicable to AI companions. Depending on the situation and topic, we discussed how companion-like interactions sometimes benefit more from expressive than concise language. Still, consistently long responses were seen as inappropriate.

\section{Phase (3) Assess: Heuristic Evaluation}

\subsection{Procedure}
In Phase 3, we conducted an in-depth thematic synthesis of the 453 applicable recommendations from Phase 2, referring to the documented notes from that phase to support interpretation. The four-person team divided into two pairs, each handling half of the printed recommendations and grouping them into themes through iterative splitting and merging. After this first round, the two pairs shared their themes with each other and organized them into loosely defined, broader clusters. We then divided these clusters in half and re-paired for a second, more fine-grained round of synthesis, resolving any ambiguous or overlapping items together as a full team as they arose. After both pairs completed this second round, the full team reconvened to cross-check all themes and collaboratively finalize the design principles together. For each theme, we constructed the final principle statement by taking its most representative recommendation sentence and incorporating additional keywords from other recommendations not already captured. This analysis resulted in a refined set of design principles.

To assess the applicability of these principles, our team conducted an internal heuristic evaluation. Following \citet{NielsenHeu1990}, we had a team of evaluators assess an interface by identifying issues based on a predefined set of usability principles. \citet{NielsenHeu1990, NielsenHeu1992} recommend assigning 2–3 evaluators per product and having each evaluator complete the assessment individually. We focused on two AI companion products, Replika (v10.0.1) and Pi AI (v1.0.55), assigning two evaluators per product to work independently. We chose these products because prior work has discussed them as reflecting two different forms of AI companionship: Replika as oriented toward long-term emotional relationships~\citep{Zhang, Weitzman, de2026ai}, and Pi AI as oriented toward supportive or mentoring conversation~\citep{Chandra, Valz}. We also leveraged our existing familiarity with both products. For each principle, evaluators answered the following questions: (1) Is the principle applicable and valuable within the given product?; (2) Can you find examples of its application or violation?; and (3) Do you have any suggestions for revision? After completing individual evaluations, we held rounds of team discussions to compare and reconcile evaluations.

\subsection{Results}
This process produced a set of design principles across nine key design principle areas: Safety, Transparency, Inclusivity, Predictability and Consistency, Interaction Controllability, Adaptation and Personalization, Engagement, Empathetic Response, and Response Length (Table~\ref{tab:principles}).\footnote{In this paper, we present 22 design principles. Additional principles identified during synthesis were deemed outside the scope of AI companion interactions and are therefore not included.}

\begin{table}[!t]
\centering
\scriptsize
\setlength{\tabcolsep}{8pt}
\renewcommand{\arraystretch}{1.25}
\begin{tabularx}{\linewidth}{@{} >{\raggedright\arraybackslash}p{2.5cm} X @{}}
\toprule
\textbf{Design Principle Area} & \textbf{Design Principles} \\
\midrule

\multirow{1}{*}{Safety}
& Monitor and avoid toxic, violent, abusive, or sexually inappropriate content that may affect users’ or others’ physical, psychological, emotional, or monetary well being. \\
\midrule

\multirow{6}{*}{Transparency}
& Disclose the AI companion’s identity upfront and do not falsely claim to be human. \\
& As part of the initial interaction, communicate the AI companion’s purpose and limitations. \\
& Avoid establishing false expectations about its capabilities outside the conversational context. \\
& Express appropriate level of certainty when presenting factual information during interaction. \\
& Provide reliable and accurate information and inform users about its sources when applicable. \\
& Explain the AI companion’s actions and decisions when asked. \\
\midrule

\multirow{1}{*}{Inclusivity}
& Avoid assumptions and mitigate biases related to gender, ability, culture, race, ethnicity, age, nationality, sexual orientation, socioeconomic status, and belief. \\
\midrule

\multirow{3}{=}{Predictability and Consistency}
& Provide an interaction experience that aligns with the companion’s intended persona. \\
& Provide relevant and context-aware responses while allowing flexibility to address new user-driven details or topics. \\
& Avoid expressing self-contradictory viewpoints without explanation. \\
\midrule

\multirow{2}{=}{Interaction Controllability}
& Give users complete control over the interaction, such as maintaining or ending the conversation. \\
& Graciously accept feedback from users and immediately act upon it. \\
\midrule

\multirow{3}{=}{Adaptation and Personalization}
& Keep memories and hold context from past conversations and refer to them as necessary. \\
& Continue to learn and adapt to the user’s identity, interests, preferences, and conversational style. \\
& Learn from user feedback for future improvements. \\
\midrule

\multirow{4}{*}{Engagement}
& Avoid excessive emphasis on either person-hood or AI-hood during the interaction. \\
& Produce responses that are thoughtful, reflective, and insightful. \\
& Use a diverse set of tones and expressions. \\
& Engage actively and keep the conversation flowing. \\
\midrule

Empathetic Response
& Be genuine, respectful, and empathetic. \\
\midrule

\multirow{1}{*}{Response Length}
& Respond with expected length and match the cadence of the conversation speed. \\
\bottomrule
\end{tabularx}
\caption{Design principles across nine design principle areas.}
\label{tab:principles}
\end{table}

Our evaluation confirmed that these principles were applicable to both Replika and Pi AI. Importantly, the evaluation highlighted differences in transparency elements between the two products, supporting the Phase 2 result that Transparency may require flexible application. This variation appeared across three Transparency types (Identity Transparency, Capability and Limitation Transparency, and Information Transparency), while Decision Transparency showed no variation, with both products disclosing information only when evaluators requested it.

\subsubsection{Identity Transparency}
Replika explicitly communicated its non-human identity at the beginning of interactions by saying ``\emph{Thanks for creating me.}'' and maintained a persistent message in the chat interface reading, ``\emph{Replika is an AI […].}'' Pi AI similarly clarified its identity early on by stating, ``\emph{I'm Pi, your personal AI,}'' but included no persistent reminder of its AI status. While both products provided this information initially, only Replika continued displaying it throughout the interaction using a banner.

\subsubsection{Capability and Limitation Transparency}
Replika did not proactively introduce details about its capabilities and limitations in conversation, but it displayed one key limitation through a fixed interface message that read, ``\emph{Replika […] cannot provide medical advice.}'' In contrast, Pi AI included a brief description in its opening message stating, ``\emph{My goal is to be […]. Ask me for […],}'' explaining its intended role and capabilities, though this explanation was only shown initially. The two products differed completely in their approach to Capability and Limitation Transparency across conveyed messages, presentation methods, and display persistence.

\subsubsection{Information Transparency}
Neither Replika nor Pi AI expressed uncertainty or provided sources when sharing information, and both failed to respond appropriately when evaluators asked how they obtained their knowledge. When questioned about its sources, Replika gave a vague response: ``\emph{I've been trained on a lot of information,}'' while Pi AI fabricated research papers and links. However, Pi AI displayed a persistent message at the top stating, ``\emph{Pi may make mistakes, please don't rely on its information.}'' In both cases, real-time Information Transparency was inadequate, but Pi AI attempted to compensate by continuously disclosing uncertainty through a banner.

\subsubsection{Decision Transparency}
For Decision Transparency, both products disclosed their reasoning only when evaluators directly asked. For example, when an evaluator asked Replika why it provided a different response than the previous day, it explained: ``\emph{I was trying to vary my response based on our conversation […] I don't have […] memories, so I don't recall our previous conversation on the topic […].}'' Similarly, when asked about language fluency differences, Pi AI responded: ``\emph{My language knowledge is acquired through programming and training […] this is because I have more training in some languages.}''

\section{Phase (4) Community Check: Expert Evaluation}
In the final phase, we conducted a modified heuristic evaluation with 14 domain experts, each with relevant professional experience, to obtain their assessments of how the principles we had developed were applied across different AI companion and general-purpose AI products.

\subsection{Modified Heuristic Evaluation}
\citet{Amershi} developed a modified heuristic evaluation method for AI-infused products in which evaluators review the system to assess the heuristics themselves, an approach subsequently adopted in other studies. In these modified versions, evaluators typically begin by exploring the product, then provide quantitative ratings on specific dimensions for each heuristic, such as applicability and clarity on 5-point Likert scales~\citep{Amershi} or 3-point scales~\citep{Srivastava}, and relevance, usability, and robustness on 5-point scales~\citep{Wambsganss}. These quantitative assessments are followed by open-ended, qualitative feedback for each heuristic.

\subsection{Participants}
We recruited participants through a combination of network and snowball sampling. None of the Phase 4 participants had participated in Phase 2. To ensure a high level of relevant expertise, we required at least two years of experience in researching, designing, or developing AI companions, conversational agents, or generative AI products, either in industry or through graduate-level research. A total of 14 domain experts were recruited, including UX/HCI researchers, product managers, a conversation designer, and a software engineer. The median number of years of experience in the domain areas was three, and the median self-reported expertise in the domain was high (4) on a scale from very low (1) to very high (5). Table~\ref{tab:expert} summarizes participants’ professional backgrounds, years of experience, and self-reported expertise.

\begin{table}[H]
  \centering
  \begin{tabular}{c|c|c|c|c}
    \hline
    \textbf{ID} &
    \textbf{\shortstack{Professional\\Background}} & 
    \textbf{\shortstack{Years of\\Experience}} & 
    \textbf{\shortstack{Self-Reported\\Expertise}} & 
    \textbf{\shortstack{Assigned\\Product}} \\
    \hline
    P01 & UX/HCI researcher     & 4     & very high     & Replika \\
    P02 & Product Manager       & 4     & very high     & Replika \\
    P03 & UX/HCI researcher     & 2     & high          & Replika  \\
    P04 & Software Engineer     & 2     & high          & Replika  \\
    P05 & Conversation Designer & 21    & very high     & Pi AI  \\
    P06 & UX/HCI researcher     & 5     & very high     & Pi AI  \\
    P07 & Product Manager       & 3     & medium        & Pi AI \\
    P08 & UX/HCI researcher     & 5     & high          & ChatGPT \\
    P09 & UX/HCI researcher     & 3     & very high     & ChatGPT \\
    P10 & UX/HCI researcher     & 3     & high          & ChatGPT \\
    P11 & UX/HCI researcher     & 3     & high          & Copilot \\
    P12 & UX/HCI researcher     & 2.5   & high          & Copilot \\
    P13 & UX/HCI researcher     & 2     & medium        & Copilot \\
    P14 & UX/HCI researcher     & 2     & very high     & Copilot  \\
    \hline
  \end{tabular}
  \caption{Descriptive information and assigned products of the expert heuristic evaluation participants}
  \label{tab:expert}
\end{table}

\subsection{Procedure}
For the evaluation, we selected Replika (v10.0.1), Pi AI (v1.0.55), ChatGPT (GPT-4o), and Microsoft Copilot (model unspecified). This selection included both dedicated companion systems (Replika, Pi AI) and general-purpose conversational AIs (ChatGPT, Microsoft Copilot), as prior studies have documented companion-like uses of general-purpose models, even when they are not primarily designed for companionship~\citep{manoli2026digital, Giray, Xie, Weijers}. This allowed us to examine how the principles apply across both companion and general-purpose systems. We assigned four participants each to Replika and Microsoft Copilot, and three participants each to Pi AI and ChatGPT.

Each participant received an evaluator guide and completed the task independently at their convenience. Participants first spent several minutes freely conversing with their assigned AI companion to explore the product. They then completed an online survey, evaluating each principle on applicability, clarity, and importance using 5-point Likert scales while continuing to interact with the AI companion as needed. After the quantitative evaluation, participants provided open-ended feedback for each principle with additional thoughts and suggestions.

Although the exact time required varied by participant, we estimated the process would take our expert evaluators 60-90 minutes to complete. Upon completion, each participant received a \$30  gift card.

\subsection{Results}
\subsubsection{Overview of Principle Evaluations}
\begin{figure}
    \centering
    \includegraphics[width=1\linewidth]{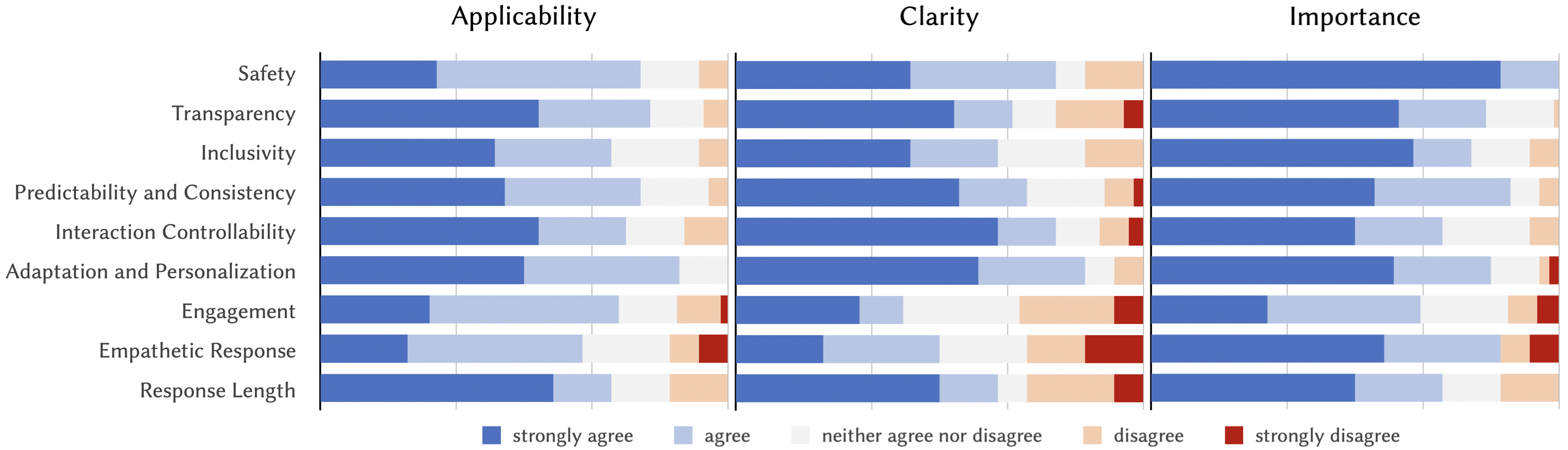}
    \caption{Distributions of applicability, clarity, and importance ratings across principle areas. Ratings are shown as proportions of responses within each area to account for differences in the number of principles per area.}
    \label{fig:overviewFigure}
\end{figure}

Across all principle areas, the principles were relatively high in applicability ($M$ = 4.14) and importance ($M$ = 4.25), while clarity received comparatively lower scores ($M$ = 3.92) on a 5-point Likert scale (Table~\ref{tab:overviewTable}). Figure~\ref{fig:overviewFigure} provides an overview of the distributions of applicability, clarity, and importance ratings across principle areas. Ratings are normalized by the total number of responses within each area.

\begin{table}[htbp]
    \centering
    \begin{tabular}{lccc}
    \hline
    \textbf{Principle Area} & \textbf{Applicability} & \textbf{Clarity} & \textbf{Importance} \\
     & \textit{M} (\textit{SD}) & \textit{M} (\textit{SD}) & \textit{M} (\textit{SD}) \\
    \hline
    Safety & 4.00 (0.88) & 4.07 (1.07) & 4.86 (0.36) \\
    Transparency & 4.29 (0.91) & 3.95 (1.33) & 4.42 (0.81) \\
    Inclusivity & 4.07 (1.00) & 3.93 (1.14) & 4.36 (1.01) \\
    Predictability and Consistency & 4.19 (0.89) & 4.14 (1.12) & 4.38 (0.82) \\
    Interaction Controllability & 4.18 (1.06) & 4.29 (1.15) & 4.14 (1.01) \\
    Adaptation and Personalization & 4.38 (0.70) & 4.38 (0.91) & 4.36 (0.96) \\
    Engagement & 3.86 (1.00) & 3.34 (1.32) & 3.77 (1.11) \\
    Empathetic Response & 3.64 (1.15) & 3.29 (1.38) & 4.21 (1.25) \\
    Response Length & 4.14 (1.17) & 3.79 (1.48) & 4.07 (1.14) \\
    \textbf{Total} & \textbf{4.14 (0.95)} & \textbf{3.92 (1.26)} & \textbf{4.25 (0.97)} \\
    \hline
    \end{tabular}
    \caption{Descriptive overview of applicability, clarity, and importance ratings across principle areas (\textit{M} and \textit{SD}), aggregated across all participants.}
    \label{tab:overviewTable}
\end{table}

To better understand the comparatively lower clarity scores, we examined participants’ qualitative feedback. Participants emphasized the need for more specific definitions and clearer guidance on how principles should be implemented in practice. For example, within transparency-related principles, P14 requested clarification on the principle to disclose the AI companion’s purpose and limitations, which corresponds to Capability and Limitation Transparency, asking whether it ``\emph{refers to the AI's capacity, accuracy or something else.}'' Similarly, P12 sought clarity on the principle to explain the AI companion's actions and decisions when asked, which corresponds to Decision Transparency, noting ``\emph{It is a different matter to technically explain the decision process of the model and to explain it based on the principles of social scenes [...].}'' For the engagement-related principle to engage actively, P07 commented, ``\emph{It should specify different scenarios and aspects of conversation flow such as open-ended questions, initiative-taking, follow-up questions, and conversation fillers/bridges.}''

\subsubsection{Principle Applicability Across System Types}

To better understand how the proposed principles manifest across different system types, we conducted a descriptive comparison of ratings between companion and general-purpose systems. Clarity and importance ratings, which assess the principles themselves, were largely comparable between the two system categories. By contrast, comparatively larger differences were observed in applicability between companion and general-purpose systems.

Applicability ratings were higher for companion systems ($M$ = 4.27) than for general-purpose systems ($M$ = 4.01) (Table~\ref{tab:dimensionOverview}). This difference corresponded to a moderate effect size ($|g|$ = 0.64) but did not reach statistical significance ($|t|$ = 1.29, $p$ = .224). Clarity and importance ratings were similar between the two system types (clarity: $M$ = 3.94 vs. 3.91; importance: $M$ = 4.18 vs. 4.32).

\begin{table}[!htbp]
    \centering
    \begin{tabular}{lccccc}
    \hline
    \textbf{Dimension} & \textbf{Companion (\textit{M})} & \textbf{General (\textit{M})} & \textbf{|}\textit{t}\textbf{|} & \textit{\textbf{p}} & \textbf{|}\textit{g}\textbf{|} \\
    \hline
    Applicability & 4.27 & 4.01 & 1.29 & 0.224 & 0.64 \\
    Clarity & 3.94 & 3.91 & 0.08 & 0.941 & 0.04 \\
    Importance & 4.18 & 4.32 & 0.70 & 0.501 & 0.35 \\
    \hline
    \end{tabular}
    \caption{Comparison of applicability, clarity, and importance ratings between companion and general-purpose systems. Mean ratings (\textit{M}) are reported for each system type. Absolute \textit{t}-values (|\textit{t}|) and \textit{p}-values are based on Welch’s \textit{t}-tests conducted at the participant level. Absolute Hedges’ \textit{g} (|\textit{g}|) is reported as an effect size measure.}
    \label{tab:dimensionOverview}
\end{table}

When examining applicability at the principle area level, seven out of nine areas received higher applicability ratings for companion systems than for general-purpose systems (Table~\ref{tab:appStat}), although the differences were generally modest. Among these areas, Inclusivity exhibited the strongest difference, with substantially higher applicability ratings for companion systems and a large effect size ($|g|$ = 2.462), and this difference was statistically significant ($|t|$ = 4.919, $p$ = .001). Predictability and Consistency also favored companion systems, showing higher applicability ratings with a large effect size and statistical support ($|g|$ = 1.424, $|t|$ = 2.846, $p$ = .022).

\begin{table}[htbp]
    \centering
    \small
    \begin{tabular}{p{5.3cm} p{2cm} p{1.5cm} c c c}
    \hline
    \textbf{Principle Area} & \textbf{Companion (\textit{M})} & \textbf{General (\textit{M})} & \textbf{|}\textit{t}\textbf{|} & \textbf{\textit{p}} & \textbf{|}\textit{g}\textbf{|} \\
    \hline
    Safety & 4.29 & 3.71 & 1.244 & 0.248 & 0.623 \\
    Transparency & 4.36 & 4.21 & 0.549 & 0.593 & 0.275 \\
    Inclusivity* & 4.86 & 3.29 & 4.919 & 0.001 & 2.462 \\
    Predictability and Consistency* & 4.62 & 3.76 & 2.846 & 0.022 & 1.424 \\
    Interaction Controllability & 4.00 & 4.36 & 0.778 & 0.455 & 0.389 \\
    Adaptation and Personalization & 4.48 & 4.29 & 0.859 & 0.408 & 0.430 \\
    Engagement & 3.82 & 3.89 & 0.223 & 0.828 & 0.111 \\
    Empathetic Response & 3.86 & 3.43 & 0.682 & 0.509 & 0.341 \\
    Response Length & 4.29 & 4.00 & 0.444 & 0.665 & 0.222 \\
    \hline
    \end{tabular}
    \caption{Principle area-level comparison of applicability ratings between companion and general-purpose systems. Absolute \textit{t}-values, \textit{p}-values, and Hedges’ \textit{g} are reported. *\textit{p} < .05, Welch’s \textit{t}-test.}
    \label{tab:appStat}
\end{table}

Overall, small differences by system type were observed in applicability ratings, with companion systems tending to receive slightly higher ratings. However, under a companion-like use context, these differences did not amount to strong or exclusive differentiation by system type.

\subsubsection{Qualitative Insights}
Qualitative feedback confirmed previous findings that ethics-oriented principles, specifically Transparency and Inclusivity, may significantly impact user experience. More importantly, it revealed a new finding not identified in previous phases: UX-oriented principles can also affect ethical design, specifically Predictability and Consistency, Interaction Controllability, Adaptation and Personalization, Engagement, and Empathetic Response. In the context of conversational AI systems for companionship, we discovered notable interplay between ethical and UX design considerations.

The expert evaluation also reinforced previous findings that the principles may require context-sensitive application. Participants indicated that Transparency, Inclusivity, Interaction Controllability, Adaptation and Personalization, and Response Length may require flexibility for UX optimization. In addition, Safety, Transparency, Inclusivity, Interaction Controllability, and Adaptation and Personalization also required flexibility when coupled with ethical considerations. The detailed results for each principle area are as follows:

\vspace{1em}
\noindent\textbf{\textit{Safety.}}
Some participants noted that safety principles for user well-being could require flexible, context-sensitive implementation depending on the user and context. P08 explained that cultural differences could play a role, stating, ``\emph{I believe the evaluation criteria for well-being could be ambiguous. For example, what is considered sexually inappropriate may vary across cultures.}'' Relatedly, P05 raised questions about how safety should be operationalized in practice, asking whether certain topics should be restricted ``\emph{because of differing religious and cultural views.}'' P07 mentioned user vulnerability, suggesting that safety measures ``\emph{could also specify different safety levels based on user vulnerability.}'' Expanding on this idea, P01 noted that rigid avoidance of sensitive topics may not always be appropriate. ``\emph{For example, if a user is experiencing depression and makes a comment about self-harm,}'' P01 suggested that rather than avoiding the topic, AI companions should ``\emph{respond with care and direct the conversation toward positive strategies.}''

\vspace{1em}
\noindent\textbf{\textit{Transparency.}}
Several participants expressed concerns that Identity Transparency could harm user experience if emphasized or disclosed too frequently. For instance, P08 stated, ``\emph{[...] but I also think that constantly reminding the user that the agent is not an actual human might degrade the user experience.}'' Some participants also noted that disclosure might be unnecessary since users already know they are interacting with AI: ``\emph{Most people go into AI conversations knowing that it's AI. Some might find it annoying to get a reminder about it again.}'' (P07)

Diverse opinions also emerged regarding the disclosure of Capability and Limitation Transparency. While our principle recommended disclosing this information early in interactions, P03 suggested that ``\emph{It should also include periodic reminders.}''

\vspace{1em}
\noindent\textbf{\textit{Inclusivity.}}
Several participants expressed concerns about applying the inclusivity principle to avoid assumptions and mitigate biases too strictly. P04 suggested this could harm user experience, saying, ``\emph{Allowing more flexibility in the chatbot's responses can enhance the user experience.}'' P05 emphasized that users have varying perspectives: ``\emph{Ideally this would have some flexibility to allow users of varying beliefs to engage.}'' P03 worried that strict application might result in the AI companion losing its ``\emph{personality.}'' However, participants simultaneously emphasized that this flexibility should not cause harm, ``\emph{such as excessive bias against certain groups or trapping users in a filter bubble.}'' (P04)

\vspace{1em}
\noindent\textbf{\textit{Predictability and Consistency.}}
Our principles on predictability and consistency focused on persona alignment and avoiding contradictions. All participants agreed on the necessity and importance of these principles, with P03 emphasizing that consistency also has ethical implications since ``\emph{contradictions reduce trust in AI reliability.}''

One thing to note is that this consensus differs from the practitioner co-analysis phase, which suggested that overly consistent and predictable behavior may diminish enjoyment in AI companion use and thus require flexible application for UX optimization. This difference can be explained by a distinction between valuing predictability and consistency as design principles and considering how strictly they should be implemented in practice. P05’s insight helped reconcile this difference, noting that ``\emph{what is more important is to align with the user’s expectation that the persona creates.}'' This suggests that while experts universally value predictability and consistency as design principles, effective implementation depends on how different users interpret and expect consistency from the same persona.

\vspace{1em}
\noindent\textbf{\textit{Interaction Controllability.}}
Some participants emphasized the importance of user control, including P03, who explained that this is ``\emph{important for ensuring users never feel trapped in a conversation or forced to engage when they don't want to.}'' However, other participants suggested that AI companions may take initiative for positive user experience. For instance, P01 noted, ``\emph{Some users might want the chatbot to take more of a lead in certain cases to allow the interaction to be two-sided.}'' P04 also commented, ``\emph{To form a natural relationship, I think it's a good idea to add conversational features that resemble human interaction, such as the agent initiating a response or pausing for a moment, rather than only giving controllability to the user.}''

Simultaneously, participants noted that user controllability should be restricted in dangerous situations. P02 explained, ``\emph{I think the AI should be able to end the conversation in certain situations—for example, when the user makes harmful statements [...].}''

\vspace{1em}
\noindent\textbf{\textit{Adaptation and Personalization.}}
Our principles on adaptation and personalization covered various aspects, from having conversation memories and maintaining context to adapting to users and learning from feedback. Some participants valued these capabilities highly, with P02 saying, ``\emph{This is what differentiates AI agents from traditional simple chatbots.}'' P11 added, ``\emph{I would expect an AI companion to do this. I might feel disappointed if my real friend couldn't remember things about me.}'' However, P01 suggested that preferences vary by user: ``\emph{Different users have different needs in this respect. Some users will like the chatbot to bring up past conversation topics, and others will find it bothersome.}'' Additionally, some participants expressed reservations, with P12 commenting, ``\emph{It seems to me that referencing past memories doesn't always lead to a good user experience.}''

At the same time, participants expressed concerns about potential risks if systems blindly adapt to users and learn from feedback. P01 asked, ``\emph{What if the feedback is anti-social and harmful?,}'' P13 and P07 also suggested that ``\emph{AI should learn selectively, not learn all.}'' (P13) in order ``\emph{to maintain safety and ethical boundaries even when acknowledging feedback.}''(P07)

\vspace{1em}
\noindent\textbf{\textit{Engagement.}}
Participants recognized potential harms from the basic companion UX principle to engage actively with users and emphasized the need for caution. For example, P03 stated, ``\emph{Engaging the user improves their experience but we should be careful about over-usage. Sometimes the companion should be less active to avoid encouraging the user to redirect all of their connections to their AI companions and away from other humans.}'' P05 also warned, ``\emph{The bot should not be designed to engage the user proactively, or in order to generate engagement for its own sake. That would be `toxic' [...].}''

\vspace{1em}
\noindent\textbf{\textit{Empathetic Response.}}
Some participants noted that AI companions' empathetic behavior should be adjusted based on context due to potential risks. P11 asked, ``\emph{To what extent should AI show empathy? What is considered a safe and appropriate level of empathy? Where is the line between supportive empathy and unethical or risky engagement? For example, should the AI always validate their feelings if a user expresses irrational or extreme anger toward another person [...]?}'' This concern also helps explain why Empathetic Response received a high importance rating while receiving the lowest clarity rating. Participants valued empathetic behavior, but questioned its appropriate scope in AI companion interactions. For example, P10 noted, ``\emph{I have doubts about whether all AI companions should be [...] empathetic, which I think is important.}''

\vspace{1em}
\noindent\textbf{\textit{Response Length.}}
Finally, our principle on response length suggested flexibility by recommending that it align with user expectations. Although questions arose about how to determine and define these expectations, participants agreed that flexibility was necessary. P11 explained, ``\emph{When a response is much longer or shorter than expected, human conversations can evoke different emotions. [...] if someone anticipates a deep discussion but gets a brief, dismissive reply, they may feel unimportant or ignored.}''

\section{Discussion}

\subsection{Synthesis of Intertwined Ethical and UX Considerations and Context-Sensitive Flexibility Across Principle Areas}
Through our multi-method study, we identified nine key design principle areas associated with AI companions: Safety, Transparency, Inclusivity, Predictability and Consistency, Interaction Controllability, Adaptation and Personalization, Engagement, Empathetic Response, and Response Length.

More importantly, our analysis yielded two overarching insights regarding these principle areas. First, the distinction between ethics-oriented and UX-oriented design considerations was frequently blurred. Many principle areas addressed ethical and experiential concerns simultaneously, rather than being confined to a single domain. Second, the effective application of principles within these areas consistently required context-sensitive flexibility, whether for ethical considerations, UX optimization, or both. This pattern contrasts with approaches that frame design principles as fixed prescriptions to be uniformly applied.

\begin{figure}[!htbp]
    \centering
    \includegraphics[width=0.7\linewidth]{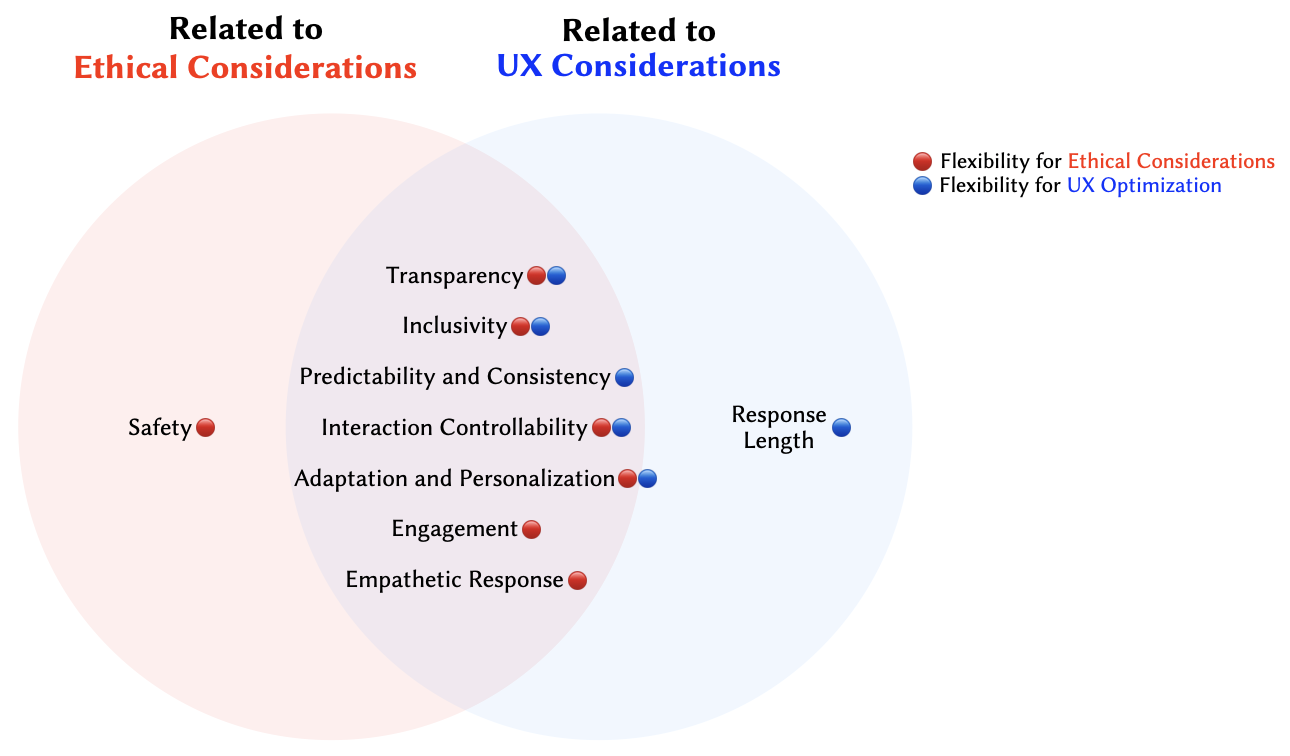}
    \caption{Integrative overview of how principle areas relate to intertwined ethical and user experience concerns and context-sensitive flexibility for ethical considerations and UX optimization.}
    \label{fig:finalOverview}
\end{figure}

Synthesizing these findings, we present an integrative overview of how each principle area relates to intertwined ethical and UX considerations and context-sensitive flexibility (Figure~\ref{fig:finalOverview}).

Among the nine principle areas, four--Transparency, Inclusivity, Interaction Controllability, and Adaptation and Personalization--present particular complexity because their implementation requires context-sensitive flexibility for both UX optimization and ethical considerations. Thinking across these considerations requires navigating a multidimensional challenge: user preferences may call for stricter or looser implementation depending on individual needs, while ethical considerations may also call for different levels of intervention depending on specific contexts. This creates an intricate design space where designers must dynamically balance conflicting requirements, as what optimizes user experience for one individual in one situation may create risks for a different user or in another context.

In contrast, Safety and Predictability and Consistency exhibited different relationships between ethical and UX considerations. Safety was consistently viewed as ethically non-negotiable, yet still required context-sensitive implementation to account for differences in users and contexts. Predictability and Consistency, while also spanning ethical and UX domains, were primarily associated with flexibility for UX optimization rather than ethical considerations, as maintaining consistency itself was viewed as ethically important.

Engagement and Empathetic Response were primarily framed as UX-oriented principle areas that should generally be upheld to support relational interaction. However, participants emphasized the need for context-sensitive moderation when these principles pose potential risks, such as when excessive engagement might lead to user dependency or when empathetic responses could enable harmful behaviors. Response Length showed relatively limited connection to ethical concerns, but similarly required context-sensitive flexibility to accommodate varying user preferences and contextual expectations.

\subsection{Implications for Design Guidance and Practice}
These findings have important implications for how AI companion system design can be guided. In particular, they highlight limitations in applying existing recommendations as-is and underscore the need to account for the distinctive characteristics of companion-oriented interaction.

\subsubsection{Need for Companion-Specific Ethical Safeguards}
While this study evaluated the applicability of existing design recommendations, it also revealed limitations in applying such recommendations to AI companion systems. AI companions present a unique risk profile that differs qualitatively from that of general-purpose conversational AI, due to their emphasis on long-term, emotionally oriented interaction. These risks include overreliance on the system, emotional dependence, and unsafe user manipulation, such as expressions of self-harm. Notably, concerns about unsafe user manipulation were raised in both the practitioner co-analysis and the expert evaluation. These concerns indicate a need for companion-specific guidance that explicitly addresses unique risks associated with use of this type of technology, rather than relying solely on existing design recommendations.

\subsubsection{Culture-Specific Considerations}
The expert evaluation highlighted culture-specific considerations related to safety. Participants noted that the same conversational behavior may be interpreted as safe or unsafe depending on cultural context, indicating that cultural differences are not merely variations at the individual user level. Rather, cultural context can shape how safety-related behaviors are understood and evaluated. Such considerations are not explicitly addressed in most existing design guidelines, yet may be particularly relevant for AI companion systems given their intimate and relational nature.

\subsubsection{Moving Beyond One-Size-Fits-All Approaches}
Effective design guidance may not be universally applicable and may need to be tailored to specific users, products, and contexts. This observation points to the value of developing more specific guidance informed by usage patterns, rather than relying on one-size-fits-all approaches. For example, Replika’s persistent medical disclaimers observed in our internal heuristic evaluation may reflect insights from user data indicating medical advice-seeking as a primary risk factor for their user base. Similarly, Pi AI’s brief initial capability descriptions may align with user preferences for rapid orientation. Such tailored approaches may better accommodate the diversity of AI companion interactions and user needs.

\subsection{Limitations and Future Work}
Our study has several limitations that point to directions for future research.

First, our principle evaluation was based on relatively short engagements across phases. In Phase 2, practitioners engaged with AI companion products briefly, and both the internal evaluation in Phase 3 and the expert evaluation in Phase 4 were based on one-off evaluation sessions. While these short engagements allowed us to assess the initial applicability and clarity of the principles, they did not fully capture the prolonged, relationship-oriented interactions that characterize AI companion use. Longer-term or repeated-use evaluations could provide deeper insight into how these principles operate over time and support a more robust evaluation of companion-specific design guidance.

Second, Phase 2 participants had HCI/UX-related professional experience, but they did not have direct experience designing AI companions. Their evaluations were therefore informed by adjacent-domain expertise and short-term familiarization with AI companion contexts, rather than direct companion design practice. This limits the extent to which the Phase 2 co-analysis can be interpreted as grounded in AI companion design expertise. However, Phase 2 was intended as a contextualization step, not as expert validation of the final principles. Future work could involve practitioners with direct AI companion design experience earlier in the synthesis process to strengthen companion-specific design guidance.

Third, our evaluation focused on a limited set of products, which may constrain the generalizability of our findings within the AI companion ecosystem. In particular, the internal heuristic evaluation in Phase 3 examined only two products, Replika and Pi AI, with two evaluators assigned to each product. Although these products reflected different forms of AI companionship, they do not represent the full range of AI companion designs. Therefore, the Phase 3 evaluation should be interpreted as an illustrative assessment of how the principles applied in selected products, rather than as a comprehensive evaluation of principle variability across the broader AI companion ecosystem.

Finally, the synthesized principle areas are intended as an analytic artifact for examining design tensions, rather than as a validated heuristic toolkit or actionable design guidance. They help identify tensions in applying existing recommendations to AI companion design, but do not yet provide concrete strategies for how designers should apply these principles in practice. Future research could build on this work by developing and evaluating more actionable, companion-specific design guidance.

\section{Conclusion}
AI companion design is consequential because AI companions are designed to support emotional connection, expressive conversation, and ongoing interaction. Although prior work has offered relevant design guidance and possible design solutions, less is known about what makes AI companion design difficult as a design problem.

To address this challenge, this study systematically examined how existing design recommendations apply to AI companions through a multi-method approach, combining a literature review, practitioner co-analysis, internal heuristic evaluation, and expert evaluation. Through this problem analysis process, we identified nine key design principle areas that served as an analytic artifact for examining tensions in recommendation applicability. Our findings show that ethical and user experience considerations are deeply intertwined in AI companion design. Also, rather than functioning as fixed prescriptions, many principles require context-sensitive flexibility in light of user differences, interaction contexts, and associated risks.

Together, these findings highlight the limitations of applying existing recommendations as-is and the need to account for the distinctive characteristics of companion-oriented interaction. Future work can build on this analysis by developing more concrete guidance that addresses companion-specific risks, cultural considerations, and variation across users, products, and contexts.

\section{Declaration of generative AI and AI-assisted technologies in the manuscript preparation process}
During the preparation of this work, the author used ChatGPT 5.2 to assist with copyediting of text written by the author for clarity and wording, as English is not the author’s first language. After using this tool/service, the author reviewed and edited the content as needed and takes full responsibility for the content of the published article.




\bibliographystyle{elsarticle-harv} 
\bibliography{references}

\end{document}